\documentclass[12pt]{article}

\usepackage{scicite}
\usepackage{amssymb}

\usepackage{times}
\usepackage{graphicx}
\usepackage{bm}

\newenvironment{sciabstract}{%
\begin{quote} \bf}
{\end{quote}}

\newcounter{lastnote}

\title{In-Plane Spectral Weight Shift of Charge Carriers in YBa$_{2}$Cu$_{3}$O$_{6.9}$}
\author
{A.V. Boris,$^{\ast}$ N.N. Kovaleva,$^{\S}$ O.V. Dolgov,\\
T. Holden,$^{\P}$ C.T. Lin, B. Keimer, and C. Bernhard\\
\normalsize{Max-Planck-Institut f\"{u}r Festk\"{o}rperforschung,}\\
\normalsize{Heisenbergstrasse 1, D-70569 Stuttgart, Germany}\\
\normalsize{$^\ast$ To whom correspondence should be addressed. E-mail:  A.Boris@fkf.mpg.de}\\
\normalsize{$^{\S}$Also at Institute of Solid State Physics,
Russian Academy of Sciences,}\\
\normalsize{Chernogolovka, Moscow distr., 142432 Russia.}\\
\normalsize{$^{\P}$Now at Department of Physics, Brooklyn College of the City University of
New York,}\\
\normalsize{Brooklyn, NY 11210, USA.}\\
}

\date{}

\begin{document}
\baselineskip22pt
\maketitle

\begin{sciabstract}
The temperature dependent redistribution of the spectral weight
of the $\bf CuO_2$ plane derived conduction band of the $\bf
YBa_2Cu_3O_{6.9}$ high temperature superconductor ({\bf \em T}$\bf _c$ = 92.7
K) was studied with wide-band (from 0.01 to 5.6 eV) spectroscopic
ellipsometry. A superconductivity - induced transfer of the spectral weight
involving a high energy scale in excess of 1 eV was observed. 
Correspondingly, the charge carrier spectral weight was shown to decrease
in the superconducting state.
The ellipsometric data also provide detailed information about 
the evolution of the optical self-energy in the normal and superconducting
states.
\end{sciabstract}

\newpage
The mechanism of high-temperature superconductivity (HTSC) is one of the main unsolved
problems in condensed matter physics. An influential class of theories predicts that
HTSC arises from an unconventional pairing mechanism
driven by a reduction of the kinetic energy of the charge carriers in the superconducting (SC) state
\cite{Anderson,Chakravarty1,Hirsch1}. This contrasts with the conventional Bardeen-Cooper-Shrieffer
(BCS) model, where 
correlations of the charge carriers below the SC transition temperature, $T_c$,
bring about an increase in their kinetic energy \cite{Tinkham,Chakravarty2},
which is over-compensated by a reduction of the potential energy due to the phonon-mediated
attraction. Within a nearest-neighbor tight-binding model, 
measurements of the optical conductivity $\sigma _{1}(\omega )=Re[\sigma (\omega )]$
can provide experimental access to the kinetic energy $\langle K\rangle$
via the sum rule for the spectral weight
SW$(\widetilde{\Omega })=\int_{0}^{\widetilde{\Omega }}\sigma
_1(\omega)d\omega =\frac{\pi e^2 a^2}{2\hbar^2V_u}\langle -K\rangle$, where $a$ is the in-plane
lattice constant and $V_u$ the unit cell volume \cite{Norman,Hirsch1,Maldague,online}.
The upper integration limit, $\widetilde{\Omega }$, needs to be high
enough to include all transitions within the conduction
band but sufficiently low to exclude the interband transitions.
Precise optical data may thus enable one to address the issue of a kinetic
energy - driven HTSC pairing mechanism.
In fact, optical measurements have ruled out a lowering of the kinetic energy along the 
$c$ axis (perpendicular to the highly conducting CuO$_2$ planes) as the sole 
mechanism of HTSC \cite{Tsvetkov}, but have also shown that it can contribute significantly 
to the superconducting condensation energy of multilayer copper oxides \cite{Basov,Boris}.
Recently, experimental evidence for an alternative mechanism driven by a reduction
of the in-plane kinetic energy \cite{Hirsch1} has been reported \cite{Marel,Santander}.
The comprehensive data set presented here, however, demonstrates that this scenario is not viable.

We performed direct ellipsometric measurements of the complex dielectric function,
$\varepsilon (\omega) = \varepsilon_1 (\omega)+i\varepsilon_2 (\omega)=1+4\pi i\sigma(\omega)/\omega$, 
over a range of photon energies extending from the far infrared ($\hbar\omega = 0.01$ eV)
into the ultraviolet ($\hbar \omega = 5.6$ eV) \cite{online}. 
We focus here on the $a$-axis component of $\varepsilon (\omega)$ of a detwinned $\rm YBa_2Cu_3O_{6.9}$
crystal at optimum doping ($T_c=92.7 \pm 0.4$ K) \cite{online}. 
In agreement with previous reports \cite{Marel,Santander,Little,Ruebhausen},
we observed a SC-induced transfer of SW involving an unusually
high energy scale in excess of 1 eV.
However, our data provide evidence for a SC-induced
decrease of the intraband SW and are thus at variance
with models of in-plane kinetic energy - driven pairing.
Additional data along the $b$ axis of the same crystal and for slightly underdoped
$\rm Bi_2Sr_2CaCu_2O_8$ ($T_c =86 \pm 0.5$ K) supporting these conclusions 
are presented in the supporting online text and figures.

Figure, 1A and B, shows the difference spectra $\Delta \sigma _{1a}(\omega )$
and $\Delta \varepsilon_{1a}(\omega )$ for the normal and SC states
(the measured spectra are displayed in Fig. S1). Figure 2,C to F, details
the temperature dependence of $\sigma _{1a}$ and $\varepsilon _{1a}$, averaged over
different representative photon energy ranges.

First we discuss the $T$-dependent changes in the normal
state. The most important observation is the smooth 
evolution of $\Delta \sigma _{1a}(\omega )$ and $\Delta
\varepsilon _{1a}(\omega )$ over an energy range of at least
0.1 to 1.5 eV. The additional features in $\Delta \sigma _{1a}(\omega )$
and $\Delta \varepsilon _{1a}(\omega )$ above 1.5
eV arise from the $T$-dependent evolution of the interband
transitions \cite{Kircher,Cooper}. Apparently, the response below 1.5 eV is
featureless and centered at very low frequency below 50 meV. It can hence be ascribed
to a Drude peak [due to transitions within
the conduction band or a narrowly spaced set of conduction bands \cite{bilayer}]
whose tail at high energy is significantly enhanced by inelastic interaction of the charge
carriers. A narrowing of the broad Drude peak at low $T$ accounts
for the characteristic $T$-dependent SW shift from high to low energies,
while it leaves the intraband SW unaffected \cite{Maksimov}.
SW is removed from the high energy tail, which involves a surprisingly large
energy scale of more than 1.5 eV, and transferred to the ``head'' near the
origin. As a consequence, $\sigma _{1a}(\omega )$ curves at
different temperatures intersect; for instance around $\omega \sim 20$
meV for $\sigma _{1a}(\omega )$ curves at 200 and 100 K [inset of Fig. S1A].
Furthermore, as detailed in the online material,
the integration of $\Delta \sigma _{1a}(\omega )$ above the intersection point yields a SW loss that is
well balanced by the estimated SW gain below the intersection point, so that the total SW
is conserved within the experimental error. 

The $T$-dependence of $\varepsilon _{1a}$ affords an independent
and complementary way to analyze the SW shift from high to low
energies. Figure 1 shows that $\sigma _{1a}$ and
$\varepsilon _{1a}$ follow the same $T$-dependence in the normal
state; that is, a concomitant decrease of both quantities with
decreasing $T$ is observed at every energy over a wide range 
from 0.05 to 1.5 eV. A simple analysis of these data
based on the Kramers-Kronig (KK) relation \cite{KK} confirms that the
SW lost at high energies is transferred to energies below 0.1 eV.
We note that the blue-shift of the zero-crossing of $\varepsilon
_{1a}$, $\omega |_{\varepsilon_1=0}$, [inset of Fig. S1B] can be
explained by the narrowing of the
broad Drude-peak alone, without invoking a change of the total
intraband SW.

We now turn to the central issue, the evolution of
the SW in the SC state. It can be seen in Fig. 1 that the $T$-dependent decrease
of $\sigma _{1a}$ in the high energy range becomes even more pronounced
in the SC state. The data of Fig. 1,E and F, thus confirm previous reports of an
anomalous SC-induced SW decrease at high energies
\cite{Marel,Santander}. Figure 1, C and D, however, shows that this trend
continues down to at least 0.15 eV, and Fig. 1A reveals that the
difference spectra $\Delta \sigma _{1a}$ do not differ
substantially between normal and SC states in the energy range between 0.15 and
1.5 eV. In both cases, $\Delta \sigma _{1a}$ exhibits a continuous
decrease towards high energy, which levels off near 1.5 eV. As
discussed above, this behavior is characteristic of the narrowing
of a broad intraband response. Significant differences between the
responses in the normal and SC states are observed only below $\sim
0.15$ eV, where the SC-induced changes are dominated by formation of the
SC condensate. The latter effect has been extensively
discussed in the literature \cite{online,Tanner}.

Next we discuss the evolution of the real part of the
dielectric function, $\varepsilon _{1a}$, in the SC state, which again provides
complementary information about the SW redistribution. For the normal state, we have shown that the
transfer of SW from high to low energies gives rise to a
decrease of $\varepsilon _{1a}(\omega )$, as dictated
by the KK relation between $\varepsilon
_{1a}(\omega )$ and $\sigma _{1a}(\omega )$. Figure 1 shows
that this trend suddenly ceases in the SC state,
where $\varepsilon _{1a}$ does not exhibit a SC-induced anomaly
mirroring the one of $\sigma _{1a}$.
Whereas $\sigma _{1a}$ decreases precipitously below $T_{c}$, $\varepsilon _{1a}$ remains virtually
$T$-independent. This trend holds not only near $\omega |_{\varepsilon_1=0}\sim$
0.9 eV but persists over a wide energy range from 1.5 eV down to at least 0.15 eV.
The KK relation necessarily implies that the SW loss between 0.15 and 1.5 eV needs to be balanced by a corresponding
SW gain below 0.15 eV and above 1.5 eV \cite{online,KK}.  
In addition to the SW transfer to low energies due to the narrowing
of the charge carrier response, as discussed above,
the SC-induced SW change must therefore involve the shift of a significant
amount of SW from low energy ($\omega <\omega |_{\varepsilon_1=0}$)
to energies well in excess of $\omega |_{\varepsilon_1=0}$.
In contrast to the situation in the normal state, where the total
charge carrier SW is conserved within the experimental
uncertainty, this additional SC-induced shift of SW to high energies gives rise to a 
decrease of the total charge carrier SW. We emphasize that this
is a model-independent conclusion based solely on the KK
relation \cite{online,KK} between $\varepsilon _{1a}(\omega )$ and $\sigma
_{1a}(\omega )$, both of which are directly measured by ellipsometry.

The related SW changes can be quantified using the extended Drude formalism
where the real and imaginary parts of the optical self-energy, $\Sigma (\omega )$,
are represented by a frequency dependent mass-renormalization factor, $m^{\ast }(\omega
)/m_b$, and scattering rate, $\gamma(\omega)$, respectively \cite{online}.
The renormalized plasma frequency,
$$
\omega _{pl}^{\ast}(\omega)=\omega _{pl}\sqrt{\frac{m_b}{m^{\ast}(\omega)}}=
\omega\sqrt{\frac{\varepsilon_2^2(\omega)+(\varepsilon_\infty-\varepsilon_1(\omega))^2}{\varepsilon_\infty-\varepsilon_1(\omega)}},
\eqno(1)
$$
and the scattering rate,
$$\gamma(\omega)= \frac{\omega _{pl}^2}{\omega}\frac{\varepsilon_2(\omega)}{\varepsilon_2^2(\omega)
+(\varepsilon_\infty-\varepsilon_1(\omega))^2}=
\omega _{pl}^2 \times f[\varepsilon_1(\omega),\varepsilon_2(\omega)],
\eqno(2)
$$
can be derived from the ellipsometric data. The value of $\varepsilon
_{\infty } = 5 \pm 1$ is extracted from the high energy part of the
spectra \cite{uncert}. The normalized difference between normal
and SC states, $\Delta \omega _{pl}^{\ast}(\omega)/\omega
_{pl}^{\ast}(\omega)$, is displayed in Fig. 2A \cite{normdiff}.
Most notably, $\Delta \omega _{pl}^{\ast }(\omega )/\omega
_{pl}^{\ast }(\omega )$ saturates above 0.3 eV at a
finite value of $\sim $ 0.5 \% (thin red line). This finite
asymptotic value of $\Delta \omega _{pl}^{\ast }(\omega
)/\omega_{pl}^{\ast }(\omega )\equiv \Delta \omega
_{pl}/\omega_{pl}-1/2\times \Delta(m^{\ast }(\omega)/m_b)/(m^{\ast
}(\omega)/m_b)$ above 0.3 eV cannot be ascribed to an anomaly of
the mass-renormalization factor $\Delta (m^{\ast }(\omega
)/m_b)$, which should decrease to zero as a function of increasing
energy. Instead, the asymptotic value of $\Delta \omega
_{pl}^{\ast }/\omega _{pl}^{\ast }\ (\omega >0.3)$ is indicative
of a SC-induced change in the bare plasma frequency, $\omega
_{pl}$. With $\omega _{pl}=2.04\pm 0.04$ eV, as derived from
$\omega _{pl}^{\ast }(\omega )$ [inset of Fig. 2A] around 0.45 eV,
and SW$(\widetilde{\Omega}) = \omega _{pl}^{2}/8$ we thus obtain
a SC-induced loss of the intraband SW of $\Delta SW=5.2\pm 0.7\times 10^{-3}$ eV$^{2}$.

Figure 2B shows the normalized difference
$\Delta \gamma(\omega)/\gamma(\omega)\equiv
2\times \Delta \omega _{pl}/\omega_{pl}+\Delta f(\omega)/f(\omega)$,
with $\frac{\Delta \omega _{pl}}{\omega_{pl}}=0.5\%$ and $\Delta f(\omega)/f(\omega)$
derived directly from the data \cite{normdiff}. The anomaly of the scattering rate evidently
extends to very high energy, exhibiting a slow but steady decrease with increasing energy.
In contrast, the SC-induced anomaly in the mass renormalization factor
decreases rapidly and essentially vanishes above 0.3 eV. This decrease of $Re[\Sigma (\omega )]$
with an energy scale of $\sim $0.3 eV provides an upper limit for the spectrum of
excitations strongly coupled to the charge carriers.
The observed self-energy effects are in fact well reproduced by models
where the charge carriers are coupled to bosonic modes (such as spin fluctuations or phonons) with a cut-off
energy of about 0.1 eV.

This analysis confirms our conclusion that the charge carrier response
not only exhibits an anomalous narrowing but also loses SW in the SC
state. In the framework of a nearest-neighbor tight-binding
model, the observed SW loss corresponds to an increase
of the kinetic energy in the SC state. Although this trend is in line with
the standard BCS theory, as discussed in the introduction, an analysis
becomes more difficult beyond this simple approach
\cite{Chakravarty2,Norman}. For instance, correlation effects due to the strong
on-site repulsion $U$ of the charge carriers on the Cu ions strongly
influence the electronic structure. Within the Hubbard model, a
single-band picture becomes inadequate, and the integration should
include all Hubbard bands; that is, $\widetilde{\Omega }>U$
\cite{Hirsch2,Maldague}. Changes of the
intraband SW unrelated to the kinetic energy may also be associated with structural
anomalies known to occur below $T_{c}$ in a number of high-temperature superconductors,
including $\rm YBa_2Cu_3O_{6+x}$ \cite{Meingast}. Although the expected SW
changes due to the $T$ dependence of the lattice parameters are
negligible, changes in the relative position of certain ions, such as
the apical oxygen ions, may have a significant impact.
Finally, the SC-induced broad band SW transfer may be closely
related to the so-called pseudogap phenomenon, which has been reported to exist 
in the cuprate HTSCs even at optimal doping.
Within this approach, possible perturbation of the
momentum-distribution function over the conduction band due to the
coupling of charge carriers to spin fluctuations may contribute to the observed
effect. More systematic experimental and theoretical work is
needed in order to address these possibilities.

Our broad-band (0.01 to 5.6 eV) ellipsometric measurements
on $\rm YBa_2Cu_3O_{6.9}$ and\\ $\rm Bi_{2}Sr_{2}CaCu_{2}O_{8}$ 
suggest a sizable SC-induced decrease of the total
intraband SW. In the context of the nearest-neighbor tight-binding
model, this effect implies an increase of the kinetic energy in the SC
state, which is in line with the standard BCS theory. We, however, argue
that a microscopic understanding of this behavior requires 
consideration beyond this approach, including strong correlation effects.


{\bf Supporting Online Material}\\
www.sciencemag.org/cgi/content/full/304/5671/708/\\
Materials and Methods\\
Text\\
Figs. S1-S6\\
References and Notes\\

\clearpage
\begin{figure}\includegraphics*[width=130mm]{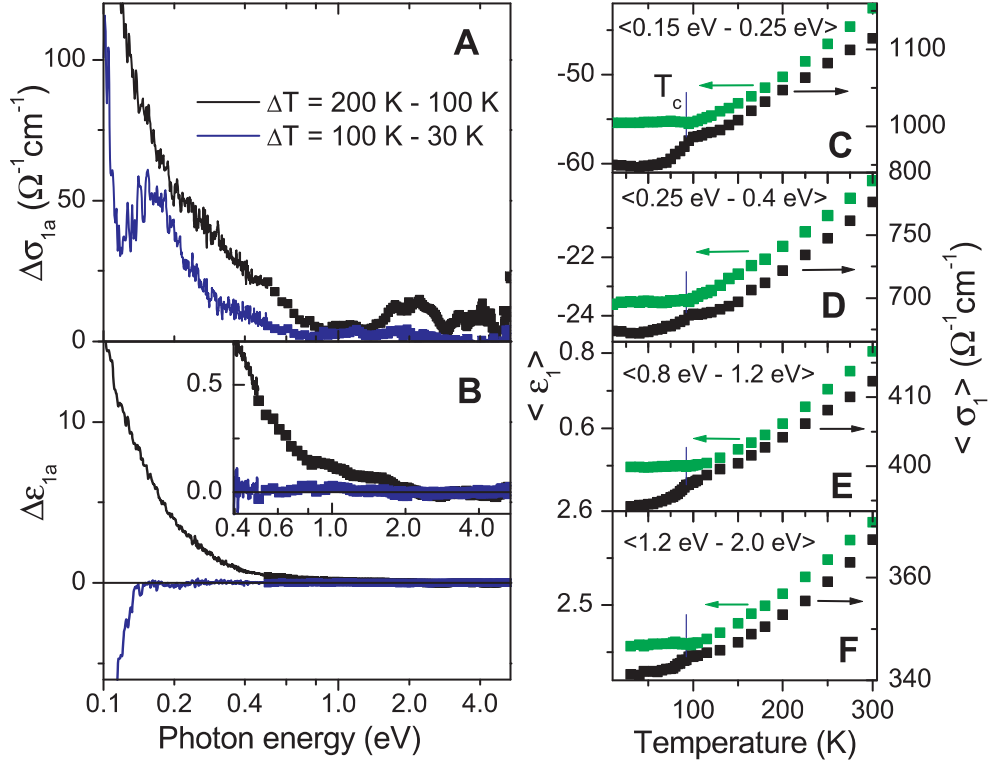}
\caption{ Difference spectra (A) $\Delta \sigma _{1a}(\omega )=\sigma_{1a}(T_2,\omega)-\sigma_{1a}(T_1,\omega)$ and
(B) $\Delta \varepsilon _{1a}(\omega )=\varepsilon_{1a}(T_2,\omega)-\varepsilon_{1a}(T_1,\omega)$ in
the normal state at $T_1=100\ K$ and $T_2=200\ K$ and below the SC transition
between $ T_1=30\ K\ (< T_c)$ and $T_2=100\ K\ (\gtrsim T_c)$.
The inset in (B) provides an enlarged view of $\Delta \varepsilon_{1a}(\omega )$
over the photon energy range from 0.4 to 4 eV.
(C-F) Temperature dependence of $\sigma_{1a}(\omega)$ (black squares)
and $\varepsilon _{1a}(\omega )$ (green squares) averaged over different energy ranges.} 
\label{Fig1}
\end{figure}

\clearpage
\begin{figure}
\includegraphics*[width=80mm]{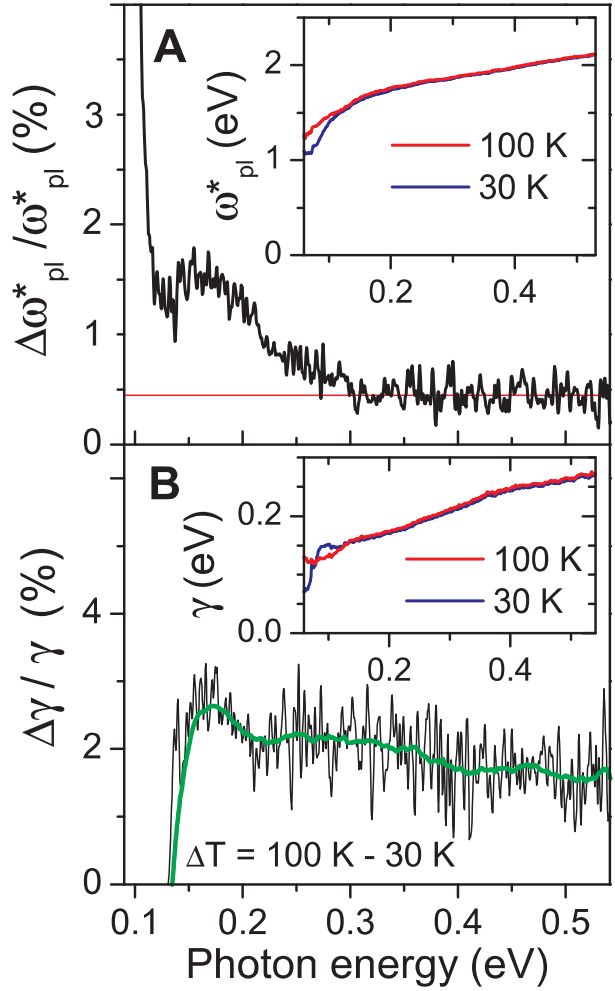}
\caption{ Normalized difference of (A) $\Delta \omega^{\ast}_{pl}(\omega)/\omega^{\ast}_{pl}(\omega)$
and (B) $\Delta \gamma(\omega)/\gamma(\omega)$, defined in the text, upon heating from 30 to 100 K. The
green curve results from smoothing of the experimental data (light black line).
The insets show (A) $\omega^{\ast}_{pl}(\omega)$ and (B) $\gamma(\omega)$
at  30 K ($<\ T_c$) and 100 K ($\gtrsim T_c$).}
\label{Fig2}
\end{figure}
\clearpage

\section*{\LARGE Supporting Online Material\\
``In-Plane Spectral Weight Shift of\\
Charge Carriers in YBa$_{2}$Cu$_{3}$O$_{6.9}$"\\
\large www.sciencemag.org/cgi/content/full/304/5671/708/}

\section*{Methods and materials}

\paragraph*{Experimental\\}

A high quality $\rm YBa_2Cu_3O_{6.9}$ single crystal grown in an
Y-stabilized ZrO crucible was annealed in flowing oxygen at 490
$^\circ$C for one week and subsequently quenched into liquid
nitrogen to achieve optimum doping. The crystal was detwinned
using a home-built apparatus that allows application of mechanical
pressure while heating in an enclosed oxygen atmosphere. Complete
detwinning was confirmed with a polarizing microscope and by x-ray
diffraction. The sample surfaces were polished to optical grade
using diamond paste. From the DC magnetization, we obtain
$T_c =92.7 \pm 0.4$ K. A slightly underdoped Bi2212 crystal with
$T_c =86 \pm 0.5$ K was grown by the traveling solvent floating
zone technique in air {\it (S1)}. For the spectral range 0.5-5.5 eV we used a
home-built ellipsometer based on a rotating analyzer as described
in Refs. {\it (S2,S3)}. For the range 0.01-0.6 eV
we used another home-built ellipsometer attached to a standard
Fast-Fourier-Transform-Interferometer (FTIR) as described in Refs.
{\it (S4,S5)}. The FIR measurements were performed at the
IR-beamline of the ANKA synchrotron light source at FZ Karlsruhe,
Germany. For the MIR measurements we used the conventional glow-bar
light source of a Bruker 113V FTIR spectrometer. Separate
measurements were performed for each of the principal axes in the
plane of incidence. The components of the dielectric tensor were
obtained using a well established numerical procedure {\it (S6-S8)}.
Ellipsometry has the advantage that it measures the complex dielectric
function directly. Its self-normalizing nature allows us to analyze the $T$-dependent
redistribution of the charge carrier SW in the normal and SC states
with high accuracy and reproducibility. Error bars of the relative
($T$-dependent) changes of the dielectric function are
smaller than 0.2 \% throughout the spectral range from 50 meV to 5 eV.

\paragraph*{Spectral Weight and Sum Rules\\}

The sensitivity of optical experiments to the kinetic energy of the charge
carriers is rooted in the fundamental sum rule SW$(\infty)=\frac{\pi n_e e^2}{2m_e}$,
where $n_e$ is the total density of electrons, $m_e$ is the free electron mass,
and SW($\Omega$) = $\int_{0}^{\Omega }\sigma_1(\omega)d\omega$ is the so-called spectral weight.
An important parameter is the upper integration
limit, $\Omega$. For $\Omega \rightarrow \infty $, SW$(\Omega)$ is
conserved and independent of the interaction between electrons. A
physically meaningful restricted sum rule, SW$(\widetilde{\Omega
}) =\frac{\pi n e^2}{2m_b}= \frac{\omega _{pl}^{2}}{8}$,
with $n$ the density of the carriers in the conduction band, $m_b$
their effective band mass, and $\omega _{pl}$ the plasma
frequency, can be obtained if $\widetilde{\Omega }$ is chosen high
enough to include all intraband transitions within the conduction
band, but sufficiently low to exclude the interband transitions.
Within a nearest-neighbor tight-binding model, the charge-carrier
SW is proportional to the single-band kinetic energy, $\langle
K\rangle$, i.e. SW$(\widetilde{\Omega })= \frac{\pi e^2
a^2}{2\hbar^2V_u}\langle -K\rangle$, where $a$ is the in-plane
lattice constant, and $V_u$ the unit cell volume {\it(S9-S11)}.
Precise optical data over a sufficiently wide energy range
(with an upper cutoff beyond $\widetilde{\Omega }$) may thus enable
one to address the issue of a kinetic-energy driven HTSC pairing mechanism.

An accurate determination of the $T$-dependent intraband
SW is a difficult experimental task, because less than 1\% of the
total charge carrier SW is redistributed in the relevant
temperature range. Another complication arises from an uncertainty
in the choice of $\widetilde{\Omega }$ due to the lack of a
transparency window ($\sigma _{1}(\widetilde{\Omega })\approx 0$)
between the intra- and interband absorption. 
Specifically, in order to deduced a kinetic energy reduction below $T_c$,
one needs to check whether the SW transferred to low energies in the SC state
indeed originates from interband transitions or rather from the tail
of the normal-state intraband response. A $T$-dependent narrowing of a very
broad charge carrier response will also give rise to a SW redistribution
from high to low energies, but it leaves the intraband SW and thus the kinetic
energy unaffected {\it(S12)}.

\paragraph*{Kramers-Kronig consistency of the T-dependent changes in $\varepsilon _{1}(\omega)$ and $\sigma_{1}(\omega)$\\}

Applying the KK analysis below $T_c$ we note
that $\sigma(\omega)$ has a singularity at $\omega =0$ due to the response of the SC condensate
$
\sigma^{sc}_{singular}(\omega )= 
\frac{1}{4\pi \lambda _{L}^{2}}\left( \pi
\delta (\omega )+i/\omega \right) ,
$
where $\lambda _{L}$ is the London penetration depth.
In the SC state one then has 
$$\varepsilon_{1}(\omega)=1-\frac{1}{\lambda _{L}^{2}\omega
^{2}}-8P\int_{0^+}^{\omega}\frac{\sigma^*_{1}
(\omega')}{|\omega'^2-\omega^2|}d\omega' + 8P\int_{\omega}^{\infty}\frac{\sigma^*_{1}
(\omega')}{|\omega'^2-\omega^2|}d\omega' ,
\eqno(S1)
$$ 
where $\sigma^*_{1}(\omega')$ is a regular function.
One can apply the KK relation as in Ref. {\it(19)},
$$\varepsilon_{1}(\omega)=1-8P\int_0^{\omega}\frac{\sigma_{1}
(\omega')}{|\omega'^2-\omega^2|}d\omega' + 8P\int_{\omega}^{\infty}\frac{\sigma_{1}
(\omega')}{|\omega'^2-\omega^2|}d\omega',
\eqno(S2)
$$
assuming that $\sigma_{1}(\omega')$ contains the superconducting
$\delta $-function at $\omega'=0$ below $T_c$. This relationship between 
$\sigma_{1}$ and $\varepsilon_{1}$ can be used to obtain a conceptually model-independent
estimate of whether SW is redistributed from high to low energy or vice versa.
If $\sigma_{1}(\omega_0)$ decreases in the vicinity of $\omega_0$ whereas $\varepsilon_{1}(\omega_0)$
remains constant, this SW decrease around $\omega_0$ needs to be balanced
by a corresponding SW gain below $\omega_0$ and above $\omega_0$ according to:
$$\int_{0}^{\omega_0^-}\frac{\Delta \sigma_{1}(\omega')}{|\omega'^2-\omega_0^2|}d\omega' 
=\int_{\omega_0^+}^{\infty}\frac{\Delta \sigma_{1}(\omega')}{|\omega'^2-\omega_0^2|}d\omega'. 
\eqno(S3)
$$
The most important point is that we observe the same trend for the $T$-changes of $\sigma_{1a}$ and $\varepsilon_{1a}$
at every energy over the entire range extending from 0.15 eV up to at least 1.5 eV.
The SW loss between 0.15 eV and 1.5 eV then needs to be balanced by a corresponding
SW gain below 0.15 eV and above 1.5 eV. In other words there is necessarily a corresponding
SW gain in the interband energy range above 1.5 eV caused by a decrease of the total intraband SW. 
We note that this our conclusion does not lose a generality seeing the
SC-induced redistribution of SW within the range below 0.15 eV. 
These changes including a singular part are dominated by formation of the SC condensate
and have been extensively discussed in the literature (see, e.g., early publications
{\it(S13-S15)}).

\paragraph*{Analysis using the Extended Drude model\\}

We discuss our data in terms of the so-called extended Drude formalism
(or more general, ``memory functions" approach {\it(S16)}) where the real and imaginary
parts of the dielectric function are given by
$$
\varepsilon (\omega )=\varepsilon_1 (\omega )+i\varepsilon_2 (\omega )=
\varepsilon_\infty+\frac{i\omega^2_{pl}}{\omega}\frac{1}{\gamma(\omega)-i\omega m^*(\omega)/m_b}.
\eqno(S4)
$$
The effective mass, $m^{\ast}(\omega )$, and scattering rate, $\gamma (\omega)$, of the charge carriers
are energy dependent due to strong inelastic scattering; $\varepsilon_\infty$ represents
the screening by interband transitions.

Concerning the formal aspects of the validity of the extended Drude-model we
remark the following points:

1. Mathematically the extended Drude formalism is the representation of the
analytical properties of a complex two-particle Green function which does
not have poles in the upper half-plane of the complex frequency $\omega $.
It remains valid both in the normal and in the SC states and can provide a good
way to obtain meaningful $T$-dependent parameters.

2. The {\it optical} mass and {\it optical} scattering
rate determine the effective mass and the decay of the electron-hole ({\it
boson}) excitations, meanwhile quasiparticle properties are
determined by the {\it fermionic} quasiparticle effective mass and the scattering
rate. They are not equivalent even in the normal state at $T\neq 0$, as discussed, e.g.,
in Ref. {\it(S17)}.

3. In the SC state the current-current polarization operator $\Pi(\omega)$ is an unified expression.
Through the coherence factor and the complex self-energy of the single particle states
it contains complete information on all temperature effects, including properties
of ``paired" and ``unpaired" quasiparticles {\it(S18,S19)}.

4. In our approach we assume that the electron-pairing interactions are limited
to energies less than 0.1 eV. The observed SC-induced anomaly in the mass renormalization factor, 
$\Delta (m^*(\omega)/m_b)$, supports this our assumption, as discussed in the
manuscript. The SC-induced changes of $\gamma(\omega)$ and $\omega_{pl}\times m^*(\omega)/m_b$ in Fig.
2 are adequately described within the electron-boson coupling model with an additional decrease of the bare plasma frequency
$\omega_{pl}$. This explanation is consistent with our conclusion based
on the KK analysis of the changes in $\varepsilon _{1a}(\omega)$ and $\sigma_{1a}(\omega)$.

\section*{Supporting Results and Discussion}

\paragraph*{{\bf \em a}-axis of $\bf YBa_2Cu_3O_{6.9}$\\}

Figure S1A shows the spectra for the real part of the $a$-axis
optical conductivity, $\sigma _{1a}(\omega)$, of $\rm YBa_2Cu_3O_{6.9}$
in the range of 0.06 - 5.5 eV at 10, 100, 200, and 300 K. The data for the FIR range
(0.01 - 0.08 eV) are displayed in the inset. Figure S1B shows the
corresponding spectra for the real part of the dielectric
function, $\varepsilon _{1a}(\omega)$. The inset gives an enlarged
view of the zero-crossing of $\varepsilon _{1a}(\omega)$.
The overall features of our ellipsometric spectra agree well with
the previously reported data on $\rm YBa_2Cu_3O_{6.9}$ based on
normal-incidence reflection {\it (16,20-22)}
and
ellipsometry {\it (22-24)},
but the superior accuracy of the present data set allows a direct determination of
the temperature dependence of both $\sigma_1(\omega )$ and
$\varepsilon_1(\omega )$ up to high energies.

The SW shift in the {\it normal state}, which involves a surprisingly high
energy scale of more than 1 eV, can be understood simply due to
the $T$-dependence of the self-energy, that is, in terms of a
narrowing of the Drude-peak at low $T$ where SW is removed from
the high energy tail and transferred to the ``head'' near the
origin. As a consequence, $\sigma _{1a}(\omega )$ curves at
different temperatures intersect, for instance around $\omega \sim 20$
meV ($\sim 70$meV) for $\sigma _{1a}(\omega )$ curves at 200 and 100 K
(300 and 200 K) (inset in Fig. S1A). From the integration
of $\Delta \sigma _{1a}(\omega)=\sigma _{1a}(\omega, T_2)-\sigma _{1a}(\omega,
T_1)$ above the intersection point, we obtain a SW loss of
$0.49\pm0.05\ eV^2$ ($0.46\pm0.04\ eV^2$) between $T_1$ = 100 K and  $T_2$ = 200 K
($T_1$ = 200 K and  $T_2$ = 300 K). This loss is well balanced by the SW gain
of $0.52\pm0.08\ eV^2$ ($0.49\pm0.06\ eV^2$) at energies below the intersection point,
so that the total spectral weight is
conserved within the experimental error. For the latter estimate, we used a linear extrapolation
of our ellipsometric data to the literature DC conductivity {\it (25)},
as indicated in the inset of Fig. S1A.

The $T$-dependence of $\varepsilon _{1a}$ affords an independent
and complementary way to analyze the SW shift from high to low
energies through the KK relationship between $\sigma _{1a}$ and
$\varepsilon _{1a}$, as outlined above. A concomitant decrease of
both $\sigma _{1a}$ and $\varepsilon _{1a}$ with
decreasing $T$ is observed over a wide energy range 
from 0.05 eV to 1.5 eV confirming that the
SW lost at high energies is transferred to energies below 0.1 eV.
We note that the blue-shift of the zero-crossing of $\varepsilon
_{1a}$ [inset in Fig. S1B] can be explained by the narrowing of the
broad Drude-peak alone, without invoking a change in the total
intraband SW. From Eq. (S4), $\omega |_{\varepsilon_1=0}(T) =
(\omega_{pl}/\sqrt{\varepsilon_\infty})\times(1-O(\gamma^2(T)/\omega^2
_{pl}))$. The $T$-dependent correction factor due to strong
inelastic scattering of the charge carriers, $\gamma^2(T)/\omega^2
_{pl}\sim 10^{-2}$, accounts for the blue shift of $\omega
|_{\varepsilon_1=0}$ at low $T$ while the screened plasma
frequency, $\omega_{pl}/\sqrt{\varepsilon_\infty}$, and hence the
total SW of the Drude-peak, $\omega _{pl}^{2}/8$, remains $T$-independent.
This brings us to the important conclusion: The observation of an
unusual SW transfer, even over a very wide energy range, and a
subsequent blue-shift of $\omega |_{\varepsilon_1=0}$ do not
necessarily imply a reduction of the kinetic energy of the charge
carriers. It may also result from the narrowing of a
Drude-response whose tail extends beyond the screened plasma
frequency.

\paragraph*{{\bf \em b}-axis of $\bf YBa_2Cu_3O_{6.9}$\\}

Figure S2A shows the spectra for the real part of the
optical conductivity, $\sigma _{1b}(\omega)$, along the one-dimensional
CuO chains in Y-123,
in the range of 0.06-5.5 eV at 10, 100, 200, and 300 K. Figure S2B shows the
corresponding spectra for the real part of the dielectric
function, $\varepsilon _{1b}(\omega)$.  The SC-induced difference
spectra $\Delta \sigma _{1b}(\omega )$ and $\Delta \varepsilon_{1b}(\omega )$
are displayed in Fig. S3A. Figure S3B presents the temperature
dependence of $\sigma _{1b}$ and $\varepsilon _{1b}$ averaged over
representative photon energy range from 0.25 eV to 0.4 eV.
In contrast to the $a$-axis response, the normal state difference spectra
$\Delta \sigma _{1b}(\omega )$ (not shown) and the temperature dependence of $<\sigma _{1b}>$ 
(Fig. S3B) exhibit nonmonotonic behavior. This is a consequence of the contribution from
the Cu-O chains to the $b$-axis optical response 
of the charge carriers, which can not be described within a single band picture. 
However, the SC-induced changes in $\sigma _{1b}$ and $\varepsilon _{1b}$ are 
very similar to those observed in the $a$-axis response. 
While $\sigma _{1b}$ decreases rapidly below T$_c$, 
$\varepsilon_{1b}$ remains almost $T$-independent, or 
even exhibits an opposite trend over a broad energy range from 0.2 eV up to 2 eV. 

\paragraph*{$\bf Bi_{2}Sr_{2}CaCu_{2}O_{8}$\\}

Furthermore, an equivalent result has been obtained for a slightly underdoped Bi-2212 crystal.
Figure S4A shows the spectra for the real part of the in-plane
optical conductivity, $\sigma _{1b}(\omega)$, of underdoped Bi2212
in the range of 0.06-0.5 eV at 10, 100, 200, and 300 K. Figure S4B shows the
corresponding spectra for the real part of the dielectric
function, $\varepsilon _{1b}(\omega)$. The difference
spectra $\Delta \sigma _{1}(\omega )$ and $\Delta \varepsilon
_{1}(\omega )$ are displayed in Figure S5, A and B.
Figures S5 and S6 show that the spectral anomalies in underdoped Bi2212
have the same character as in Y123, although they are observed over a broader
temperature range from 120 K ($>T_c\approx $ 86 K) to 50 K. 

\paragraph*{Conclusions\\}

Our ellipsometric data on both Y123 ($b$-axis) and Bi2212 show that 
the overall SC-induced features agree well with our
data measured along the $a$-axis in Y123 [see Fig. 1-2 of the main text,
and Fig. S1]. Following the line of
arguments as described in the report for the $a$-axis optical response of
Y-123 we find an evidence for a SC-induced {\it decrease} of the charge carrier
SW that is partially masked by the narrowing of a broad Drude-response.

\section*{References}
S1. B. Liang, C.T. Lin, Journal of Crystal Growth {\bf 237-239}, 756 (2002).\\
S2. J. Kircher {\it et al.},
Physica C {\bf 192}, 473 (1992).\\
S3. L. Vi\~na, S. Logothetidis, and M. Cardona, Phys. Rev. B
{\bf30}, 1979 (1984).\\
S4. R. Henn {\it et al.},
Thin Solid Films {\bf 313-314} 642 (1998).\\
S5. C. Bernhard, J. Humlicek and B. Keimer, to appear in Thin Solid Films.\\
S6. J. Huml{\'{\i}}{\v{c}}ek, A. R{\"{o}}seler,
Thin Solid Films {\bf 234} 332 (1993).\\
S7. R.M.A. Azzam and N.M. Bashara, in {\it Ellipsometry and polarized light}
(Noth Holland, Amsterdam, 1977).\\
S8. A. R\"oseler, {\it Infrared Spectroscopic Ellipsometry}, (Akademie-Verlag, Berlin, 1990).
S9. J.E. Hirsch and F. Marsiglio, Phys. Rev. B {\bf 62}, 15131 (2000).\\
S10. M.R. Norman and C. P\'epin, Phys. Rev. B {\bf 66}, 100506(R) (2002).\\
S11. P.F. Maldague, Phys. Rev. B {\bf 16}, 15131 (1977).\\
S12. A.E. Karakozov, E.G. Maksimov, O.V. Dolgov, Solid State
Commun. {\bf 124}, 119 (2002).\\
S13. N.E. Bickers, D.J. Scalapino, R.T. Collins, and Z. Schlesinger, Phys. Rev. B
{\bf 42}, 67 (1990).\\
S14. R. Akis, J.P. Carbotte, and T. Timusk, Phys. Rev. B {\bf 43}, 12804 (1991).\\
S15. O.V. Dolgov, A.A. Golubov, and S.V. Shulga, Physics Lett. A {\bf 147}, 317 (1990).\\
S16. D.B. Tanner and T. Timusk, in {\it The Physical Properties
of High Temperature Superconductors III}, D.M. Ginsberg, Ed.
(World Scientific, Singapore, 1992), pp. 363-469.\\
S17. S.V. Shulga, O.V. Dolgov, and E.G. Maksimov, Physica C {\bf 178}, 266 (1991).\\
S18. P.B. Allen, Phys. Rev. B {\bf 3}, 305 (1971).\\
S19. E. Schachinger, J.P. Carbotte, and F. Marsiglio, Phys. Rev. B {\bf 56}, 2738 (1997) and references therein.\\
S20. D.N. Basov {\it et al.}, 
Phys. Rev. Lett. {\bf 77}, 4090 (1996).\\
S21. C.C. Homes {\it et al.},
Phys. Rev. B {\bf 69}, 024514 (2004).\\
S22. S.L. Cooper {\it et al.},
Phys. Rev. B {\bf 47}, 8233 (1993).\\
S23. J. Kircher {\it et al.},
Physica C {\bf 192}, 473 (1992).\\
S24. C. Bernhard {\it et al.},
Solid State Commun. {\bf 121}, 93 (2002).\\
S25. K. Segawa and Y. Ando, Phys. Rev. Lett. {\bf 86}, 4907 (2001).

\begin{figure}\includegraphics*[width=110mm]{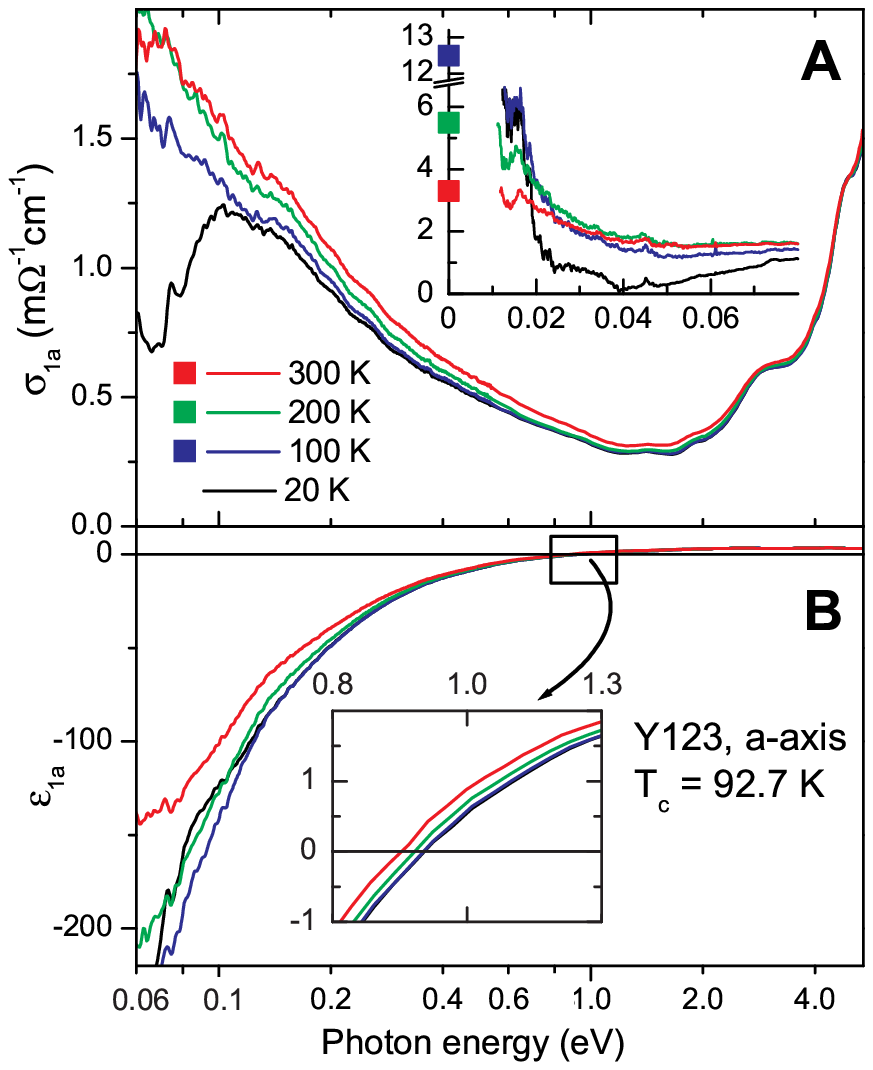}
\\
{\bf Fig. S1.} Real part of the (A) optical conductivity,
$\sigma_{1a}(\omega)$, and (B) dielectric function,
$\varepsilon_{1a}(\omega)$, along the $a$-axis of $\rm YBa_2Cu_3O_{6.9}$. The inset
to (A) shows $\sigma_{1a}(\omega)$ over the FIR spectral range.
The squares on the left ordinate represent the DC
conductivity from Ref. {\it(25)}. The inset to (B) enlarges the
spectral range around the zero-crossing of
$\varepsilon_{1a}(\omega)$.
\label{SFig1}
\end{figure}

\clearpage
\begin{figure}\includegraphics*[width=110mm]{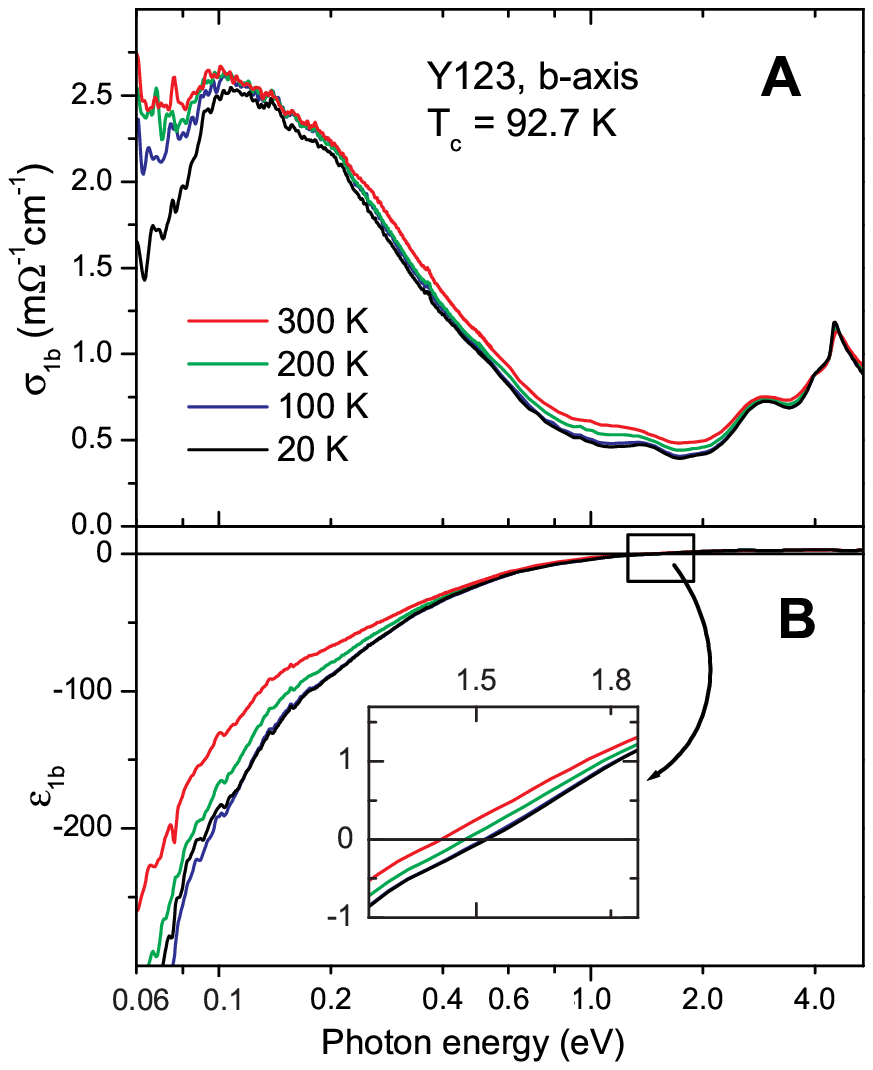}
\\
{\bf Fig. S2.} Real part of the (A) optical conductivity,
$\sigma_{1b}(\omega)$, and (B) dielectric function,
$\varepsilon_{1b}(\omega)$, along the $b$-axis of $\rm YBa_2Cu_3O_{6.9}$.
\label{SFig2}
\end{figure}

\clearpage
\begin{figure}
\includegraphics*[width=110mm]{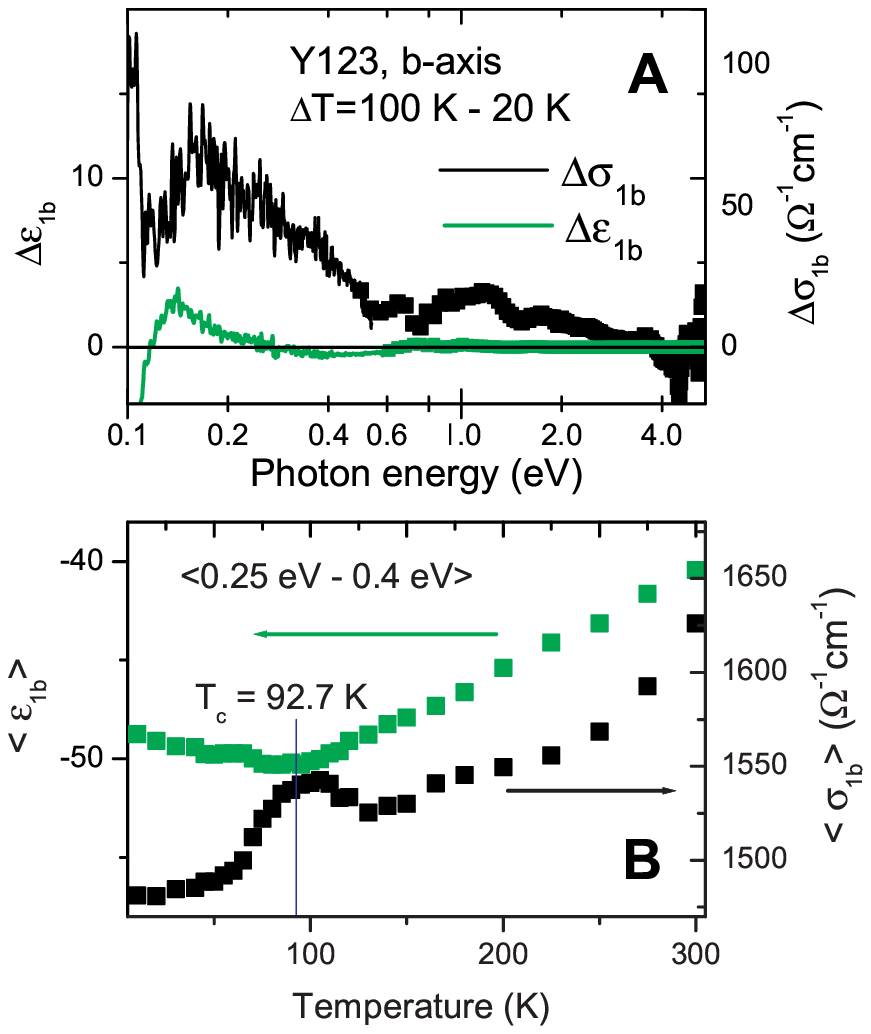}
\\
{\bf Fig. S3.} (A) Difference spectra $\Delta \sigma _{1b}(\omega )=\sigma_{1b}(T_2,\omega)-\sigma_{1b}(T_1,\omega)$ (black)
and $\Delta \varepsilon _{1b}(\omega )=\varepsilon_{1b}(T_2,\omega)-\varepsilon_{1b}(T_1,\omega)$ (green)
below the SC transition between $ T_1=20\ K\ (< T_c)$ and $T_2=100\ K\ (\gtrsim T_c)$.
(B) Temperature dependence of $\sigma_{1b}(\omega)$ (black squares)
and $\varepsilon _{1b}(\omega )$ (green squares) averaged over 0.25 - 0.4
eV energy range.
\label{SFig3}
\end{figure}

\clearpage
\begin{figure}\includegraphics*[width=110mm]{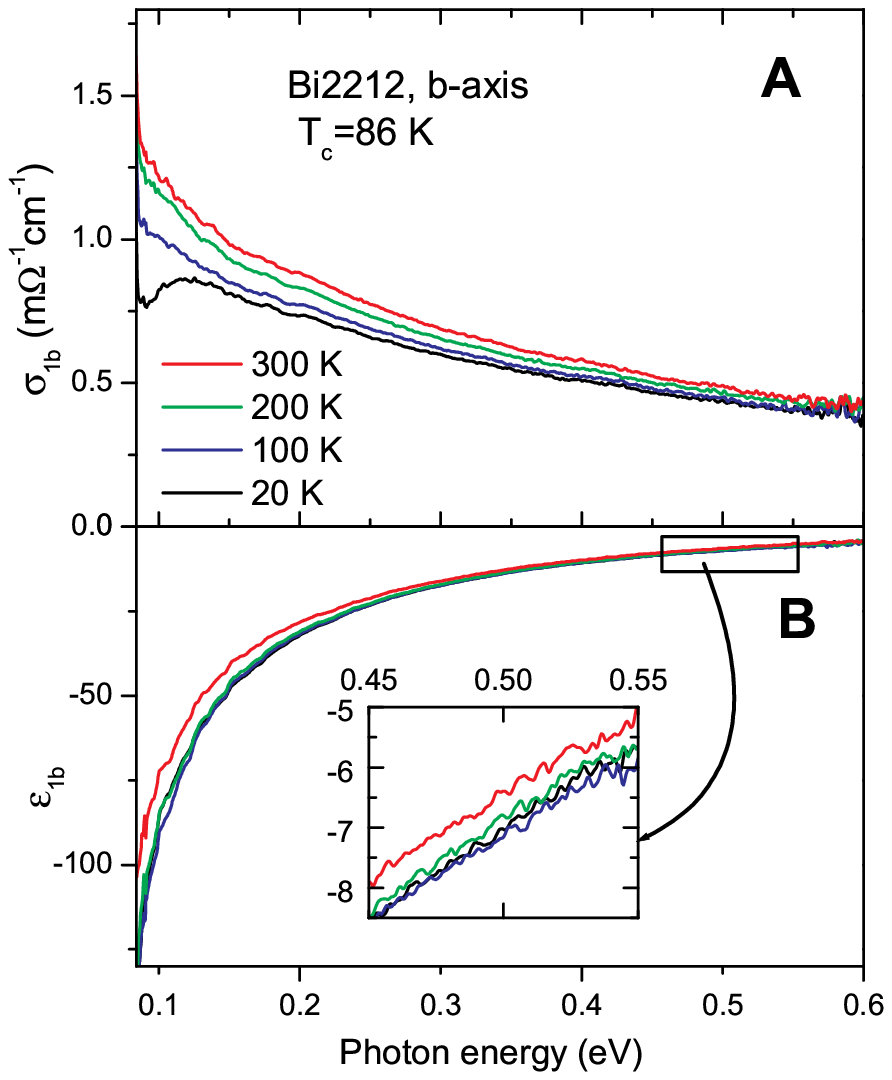}
\\
{\bf Fig. S4.} Real part of the in-plane (A) optical conductivity,
$\sigma_{1b}(\omega)$, and (B) dielectric function,
$\varepsilon_{1b}(\omega)$, of slightly underdoped $\rm Bi_2Sr_2CaCu_2O_8$
with $T_c\approx$ 86 K.
\label{SFig4}
\end{figure}

\clearpage
\begin{figure}
\includegraphics*[width=110mm]{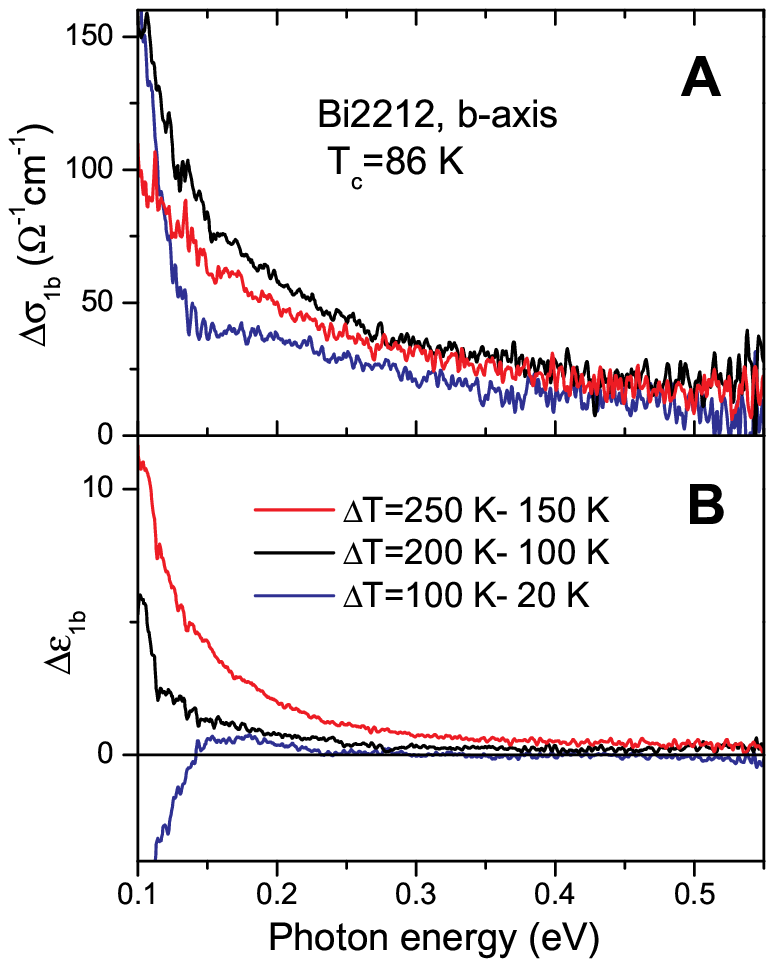}
\\
{\bf Fig. S5.} Difference spectra (A) $\Delta \sigma _{1b}(\omega )=\sigma_{1b}(T_2,\omega)-\sigma_{1b}(T_1,\omega)$ and
(B) $\Delta \varepsilon _{1b}(\omega )=\varepsilon_{1b}(T_2,\omega)-\varepsilon_{1b}(T_1,\omega)$ in
the normal state of $\rm Bi_2Sr_2CaCu_2O_8$ at $T_1=150\ K$ and $T_2=250\ K$
(red line), at $T_1=100\ K$ and $T_2=200\ K$ (black line), and between $ T_1=20\ K\ (< T_c\approx$ 86 K)
and $T_2=100\ K\ (>\ T_c)$ (blue line).
\label{SFig5}
\end{figure}
\clearpage

\begin{figure}
\includegraphics*[width=100mm]{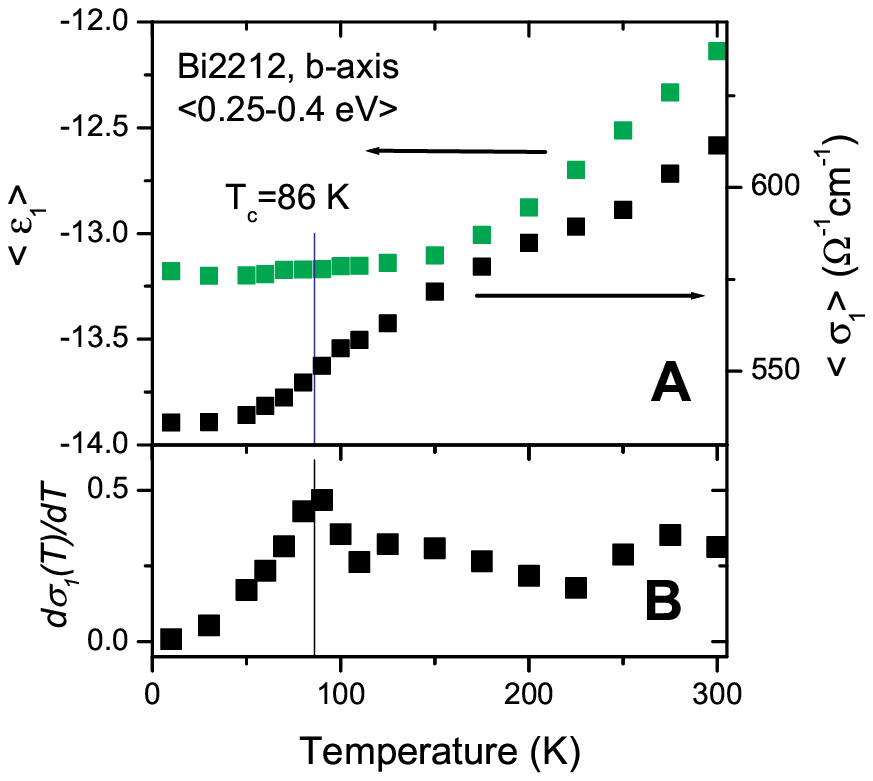}
\\
{\bf Fig. S6.} Underdoped $\rm Bi_2Sr_2CaCu_2O_8$ ($T_c\approx$ 86 K).
(A) Temperature dependence of $\sigma_{1b}(\omega)$ (black squares) and $\varepsilon _{1b}(\omega )$
(green squares) averaged over 0.25 - 0.4 eV energy range.
(B) Temperature derivative $d<\sigma_{1b}>/dT$.
\label{SFig6}
\end{figure}
\clearpage


\begin{thebibliography}{30}
\bibitem{KK}
This is most evident if the KK relations are written as\\
$\varepsilon_{1}(\omega)=1-8P\int_0^{\omega}\frac{\sigma_{1}
(\omega')}{|\omega'^2-\omega^2|}d\omega' + 8P\int_{\omega}^{\infty}\frac{\sigma_{1}
(\omega')}{|\omega'^2-\omega^2|}d\omega'
$
and combined with the $f$-sum rule  $\int_0^{\infty}\sigma_{1}
(\omega)d\omega= {\rm const}$.
\bibitem{normdiff} The normalized difference is defined as
$\Delta f(\omega)/f(\omega)=2\times\frac{f(\omega,100\ K)-f(\omega,30\ K)}{f(\omega,100\ K)+f(\omega,30\ K)}$.
\end{thebibliography}
\end{document}